\documentclass[twocolumn,showpacs,amsmath,amssymb]{revtex4}
\usepackage{bm}
\usepackage{graphicx}

\begin{document}


\title{Delocalized electrons: conductivity and superconductivity}
\author{Haigang Lu}
\email{luhg@pku.org.cn}
\affiliation{Institute of Theoretical and Computational Chemistry,
College of Chemistry, Peking University, Beijing 100871, China}

\date{\today}


\begin{abstract}
When wavefunction of crystal was projected out in atomic basis, we found that 
electrons of $s$ orbital had localization-delocalization duality, and the ones 
of $p$ and $d$ orbital were only localized in three dimensional crystal lattice. 
The existence of $s$ type delocalized electrons is based on three dimensional 
ordering of electron density in metals. Then charge carriers of metals for 
conductivity are just such delocalized electrons so that the ones for 
superconductivity are electron pairs of $s$ wave. This way could also be 
generalized to two dimensional case for CuO$_2$ plane of cuprate superconductor 
to explain its phase diagram qualitatively. We also gave some \textit{ab initio} 
explain for Hubbard model.
\end{abstract}


\maketitle


\section{Introduction}
Density functional theory(DFT) have success to predict almost all of structure 
of molecular and crystal, but band theory of DFT explains conductivity partly. 
The classical theory on conductivity is based on band structure of Bloch's 
theorem, in which charge carriers are electrons excited from valence bands to 
conduction bands or occupied in partially filled band. But it fails to explain 
the poor conductivity of Mott insulators\cite{Mott} such as CoO, MnO, CuO, which 
have half-filled band structure, while CuO is the principle component of cuprate 
high temperature superconductor. So, a unified theory of conductivity on metals 
and Mott insulators is one of keys for high temperature superconductivity.

It became practicable after we found the relations between delocalization of 
electrons and types of atomic orbitals when projecting bands of crystal to 
atomic orbitals, because delocalized electrons was just electron 
mobility\cite{ABC} with respect to their original ions in lattice. Then, charge 
carriers will be not excited electrons, but the delocalized ones near Fermi 
surface. Certainly, this way is still based on the band structure of Bloch's 
theorem. 

This paper includes five parts: First we elaborated that Slater type 
orbital(STO) was the best function to project out the atomic components from 
plane wave function, in which $\zeta$ was the best parameter to represent the 
degree of delocalization in three dimensions. Secondly, the 
localization-delocalization duality of $s$ electrons and localization of $p$ and 
$d$ ones are found, from which conductivity and Hubbard model is explained. 
Thirdly, the mechanism of superconductivity is given based on the barrier 
between localization and delocalization and the phase diagram\cite{Tallon} of 
cuprate superconductors is explained qualitatively. Fourthly, we will use 
percolation theory\cite{Stauf} to connect the microscopic and macroscopic 
mechanism of (super)conductivity. At last, we will end this paper by a 
conclusion. 

\section{Optimized pseudo atomic orbital}
\label{atomic}

To give band structure of crystal, there are two widely used methods,  plane 
waves and atomic orbitals. What we used is the former in addition to \textit{ab 
initio} pseudopotential\cite{Hamann}, and these provide a very successful scheme 
to calculate the ground state properties of crystal. But only atomic basis is 
adequate to analysis atomic properties of crystal. We have to project out the 
corresponding pseudo atomic orbitals\cite{San} from the bands of plane wave 
calculations using pseudopotential.

For posterior analysis of atomic properties, the best pseudo atomic orbital is 
single Slater type orbital 
$\chi_{nlm}(\bm{r})=N_sR_n(\zeta,r)Y_{lm}(\theta,\phi)$\cite{Slater}, where 
$N_s$ is the normalization constant, $Y_{lm}(\theta,\phi)$ is the spherical 
harmonics, $R_n(\zeta,r)=r^{n-1}\exp(-\zeta r)$ is the radical function and 
$\zeta$ is a parameter of $n$ and $l$. There are four reasons for STO: (1) 
single $\zeta$-optimised STOs can almost be equivalent to plane wave basis when 
atoms are far enough each other, (2) single STOs have the well-defined physical 
mean, where the $n$ is main quantum number, $l$ is angular momentum quantum 
number and $m$ is magnetic quantum number, (3) there is a single variable 
$\zeta$ to indicate the transformation of atomic orbitals between isolated atom 
and atom in crystal after optimised, (4) $\zeta$ is also a parameter to indicate 
the degree of delocalization of electrons with respect to the original ions, 
because the classical radius of atomic orbital is $n/\zeta$ so that the smaller 
$\zeta$ , the larger radius, the more delocalization. 

Criterion should be given to optimise the variable $\zeta$. In projecting, the 
quality of an atomic basis is quantified by its ability to represent those 
eigenstates of plane wave basis, i.e., the completeness of crystal orbitals in 
subspace spanned by the atomic basis, which is always less than 100\%. From the 
Rayleigh-Ritz minimal principle, energy of an electronic ground state 
$E=\min_\psi(\psi,H\psi)$, where $H$ is Hamiltonian and $\psi$ is a normalized 
trial function for the given number of electrons $N$. After projecting to a 
single STO of one variable $\zeta$, the above formula will become 
$E=\min_\zeta[\psi(\zeta),H\psi(\zeta)]$ so that the energy of system is a 
function of $\zeta$ and the less incompleteness of atomic basis, the lower trial 
energy of system. Then we could analyse electron state by the relation of 
incompleteness or energy and $\zeta$ of STO.

The electronic structures of supercell are calculated using \textit{ab initio} 
plane-wave code CPMD3.9.1\cite{CPMD} of density functional theory, where sizes 
of supercell are larger than 1nm$\times$1nm$\times$1nm. The exchange-correlation 
functional was Perdew-Burke-Ernzerhof\cite{PBE} ones of the generalised gradient 
approximation. The interaction of ion-electron was described by Goedecker 
pseudopotential\cite{Geodeck} with cutoff energy of plane-wave 40 or 100 Rydberg 
for $4s$ or $3d$ and $2p$ shell respectively and shell of valence electrons is 
as few as possible. The data of normal crystal structures come from experiment.


\section{localization and delocalization, Hubbard Model}
\label{local}

Fig. 1a and 1b gave some typical graphs of incompleteness of atomic basis versus 
$\zeta$ of STO in different crystal structure: $s$ type orbitals in body-centred 
cubic (K), face-centred cubic (Cu) and hexagonal close-packed (Zn) lattice, $d$ 
type ones in body-centred cubic (Cr) and face-centred cubic (Cu) lattice, $p$ 
type one in diamond (C). From these figures we could find that, there must be 
two minimums at large and small $\zeta$  separated by a group of barriers for 
electrons in $s$ orbital, and only one minimum for ones in $p$ and $d$ orbitals 
at large $\zeta$, so that electrons of $s$ orbitals have 
localization-delocalization duality, while ones of $p$ and $d$ orbitals only is 
localized in three dimensional crystal lattices. Localized electrons always 
appear in lattice and bond atoms one by one so that solids is solid, while 
delocalized electrons appear in metals to form long-range bonds so that metals 
have extensibility. By the way, the mean of "duality" here is analogous with the 
wave-particle duality in quantum mechanics for $s$ type STO of $\zeta=0$ is just 
plane wave and the one of $\zeta=\infty$ just point particle. 

\begin{figure}
\includegraphics[width=200pt]{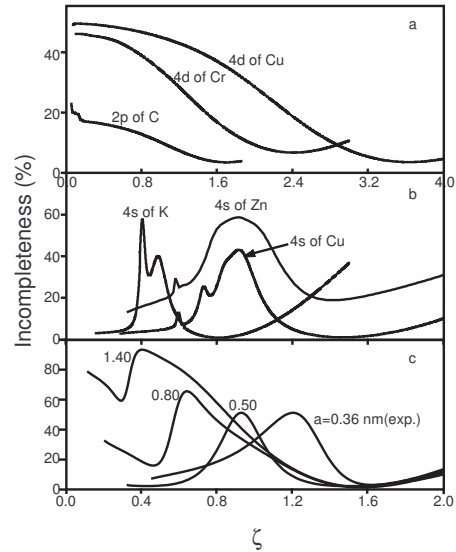}
\caption{
  \label{}Some typical graphs of incompleteness of atomic orbital 
  versus $\zeta$ of STO. $\zeta$ is in atomic unit and
  the $a$ in (c) is lattice parameter of supercell.   
  }
\end{figure}

Because localization-delocalization duality of atomic orbitals is a phenomena of 
electronic structure of crystal on \textit{ab initio} calculation, we will 
investigate the origin of minimums and barriers. Fig. 2 is graph of 
incompleteness of $4s$ orbital of Cu in four atoms cell, where the insets above 
the curve are corresponding distribution of orbitals in two dimensional section. 
In these insets, the right and left are localized and delocalized $4s$ orbitals 
respectively, which give sketch of electron distribution in crystal, and the 
middle is $4s$ orbitals in barrier, in where electron density is the lowest. 
From the view of wave, the electronic density in crystal is variable 
periodically in three dimensions so that the $s$ wave, an isotropic wave in 
three dimensions could spread from near to far away original ion, but it will 
undergo a scatter just when leaving the original ion. To $p$, $d$ and $f$ wave, 
there are two types of symmetry: $p$, $d_{z^2}$, $f_{z^3}$ waves are isotropic 
only in two dimensions so that the periods of electron density are not 
commensurable between these two dimensions and the other dimension; the others 
is anisotropic, so that the periods of electron density is not commensurable 
among all of three dimensions. Then $p$, $d$ and $f$ waves would undergo scatter 
continuously not to spread away from the localized orbitals. This localization 
of $p$, $d$ and $f$ wave in three dimensions gives one instance for Anderson 
localization of disordering\cite{Anderson}. Now we conclude that only waves 
whose symmetry is fit with medium could spread, that is to say, s wave could 
spread in three dimensional lattice and $p$, $d_{z^2}$ and $f_{z^3}$ wave could 
spread in two-dimensional sheet. 

\begin{figure}
\includegraphics[width=200pt]{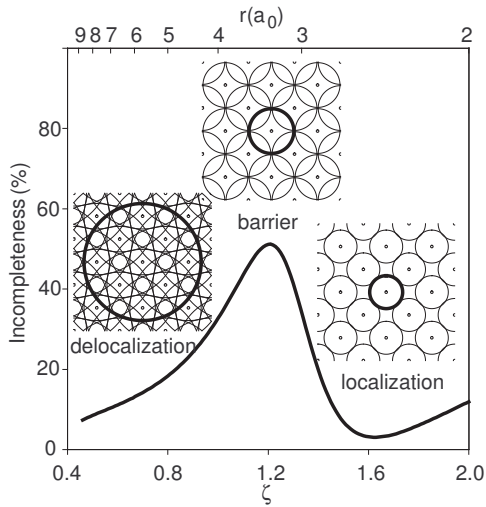}
\caption{
  \label{}Graph of incompleteness of $4s$ orbital of Cu versus
  $\zeta$ and radius of STO and corresponding distribution of 
  orbitals on two dimensional section. 
  The $a_0$ for radius is Bohr radius $0.0529$nm   
  }
\end{figure}

This relation between delocalization of electrons and types of wave has a 
far-reaching consequences that charge carriers is $s$ wave in three dimensional 
conductor such as metal conductors and superconductors, and $p$, $d$ and/or $f$ 
wave in two dimensional conductors such as graphite, organic conductors and 
cuprate superconductors. In Mott insulators such as CoO, MnO, CuO, electrons 
near Fermi surface belong to $3d$ of transition metals and $2p$ of O so that 
there are not $s$ but $d$ and/or $p$ type charge carriers. Then Mott insulators 
are insulators in three dimension. The above also explain experimental facts 
that charge carriers are $s$ waves in metal superconductors and $d$  waves in 
cuprate superconductors\cite{Mineev}, but the $d$ wave of Cu in our theory is 
not $d_{x^2-y^2}$ but $d_{z^2}$, and $2p$ of O could also be delocalized to 
contribute to charge carriers in CuO$_2$ plane and direction perpendicular to 
this plane. We conjectured that $d_{x^2-y^2}$ symmetry of charge carrier was 
just some linear combination of $2p_z$ orbitals of O in CuO$_2$ plane.

When do the delocalized electrons occur to transit Mott insulators to conductors 
or superconductors? That is to say, when the localized orbital, the fundamental 
state in bond need to be delocalized? We knew that metallic bond is unsaturated 
bond of $s$ orbitals, and $\pi$ bond in organic conductors is also unsaturated 
bond of $p$ orbitals relative  to $\sigma$ bond, so that electron deficient of 
bond is the common characteristic of delocalization of bond as well as of 
electrons. So the transition of Mott insulator-metal comes from electron 
deficient of bond, in which the doped cuprate superconductors are just some 
instances. To clarify the change of localization-delocalization duality in Mott 
metal-insulator transition, we investigate an ideal Mott transition: the 
delocalization of $4s$ of Cu for different distance of atoms in four atoms 
cell(Fig. 1c), in which more energy for delocalization of electrons is needed as 
lattice parameter become large. And this difference of energy between 
delocalization and localization is just as Hubbard energy in Hubbard 
model\cite{Mott}. The lattice parameter $a=1.4$nm means the corresponding 
``metal'' has poor conductivity for the minimum distance of atoms is large to 
$1.0$nm. And crystal of $a=0.36$nm is just true copper metal. From above we 
concluded that the distance larger, the more incompleteness of delocalization, 
the less conductivity so that metal transited to insulator.


\section{Superconductivity}
\label{conduct}


 As transforming the incompleteness of projecting to energy of system, graphs of 
incompleteness versus $\zeta$ become two electronic states separated by a energy 
barrier, and these two states can transition each other by electron tunneling. 
 When delocalized and localized electrons form bonds by electron pairing of 
 different spin, there are three possible kinds of electron pair: hybrid 
 $\alpha_L\beta_D$ and $\alpha_D\beta_L$, localized $\alpha_L\beta_L$ and 
 delocalized $\alpha_D\beta_D$ pair, where $\alpha$ and $\beta$ are spin of 
 electrons and subscripts $L$ and $D$ represent localized and delocalized state 
 respectively. In high temperature, almost all of the electrons have enough 
 energy across the barrier so that only hybrid pairs can exist. As the 
 temperature falls, some of electrons in lowest energy of states are not able to 
transit and are delocalized or localized, then delocalized and localized pairs 
 will occur because of bonding, which had been supported by two-component 
 electrons in cuprate superconductors\cite{Stevens}. It shows that the barrier 
 between localization and delocalization determines the temperature of pairing 
 of charge carriers, even transition temperature of superconductivity in metals.

 The energy barriers in metals are very low because their electron density is 
 highly homogeneous in real space so that the temperature for prohibiting 
 transition of some electron pairs should be very low, while in cuprate 
 superconductors the barriers are high because their electronic structure is 
 much more prone to inhomogeneity\cite{Emery}. One of keys of cuprate high 
 temperature superconductivity is its "phase diagram"(Fig. 3c). To explain this, 
 we should first determine the delocalized states of electrons near Fermi 
 surface as Mott insulators are doped to transit to metals (Fig. 3b), which 
 could be deduced from Fig. 1c.  The pseudogap in normal state of cuprate 
 superconductors would come from the imbalance of electrons of localized and 
 delocalized state. In high temperature, the difference of energy between 
 localized and delocalized states are negligible so that partition of electrons 
 in two states is duality and Fermi surface fill in, but as temperature falls 
 below temperature of pseudogap, this difference of energy would dominate 
 partition, then there are more localized electrons than delocalized ones and no 
 enough electrons for Fermi surface so that this surface will be broken (Fig. 
 3a)\cite{Norman}. So, pseudogap does correspond with a phase transition with 
 broken Fermi surface. On the other hand, more doping produces more itinerant 
 electrons and reduces inhomogeneity so that the barrier will decreases as 
 doping, so $T_c$ will rise first and then drop. In underdoped side, there is no 
 enough carriers and transition temperature $T_c$ is dominated by hole 
 concentration, and in overdoped side, the energy barrier is low so that $T_c$ 
 is determined by energy barrier for delocalized state.

\begin{figure}
\includegraphics[width=200pt]{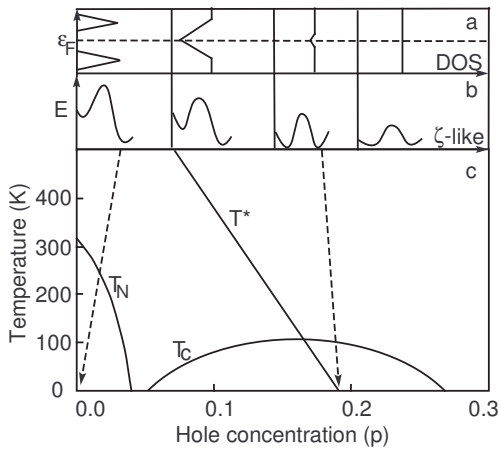}
\caption{
  \label{}Schematic phase diagram of cuprate superconductors (c)
  and corresponding schematic diagrams for density of state(DOS) 
  (a) and delocalization of orbitals (b) near Fermi surface. 
  $T_N$ is the Neel temperature for three dimensional 
  antiferromagnetic state, $T_c$ the transition temperature 
  of superconductivity and $T^*$ the temperature below which 
  the normal-state pseudogap opens, 
  $\zeta$-like is the corresponding parameter of delocalization
  in two dimensional cases.   
  }
\end{figure}


\section{Percolation}
\label{perco}

For superconductivity of cuprate superconductors, there 
is a similar critical fraction $0.05$ of hole concentration, which could be 
deduced  from the delocalized radius of electron and percolation theory. 
Percolation theory is the bridge between microscopic and macroscopic mechanism 
of (super)conductivity, in which the delocalized orbitals were thought of 
(super)conductor spheres and localized ones insulator spheres. Because the 
radius of delocalized orbital is much larger than of localized one, we should 
use the (super)conductor sphere as unit to calculate thresholds for percolation. 
Once the fraction of carriers exceeds the corresponding threshold, insulators 
will transit to metal, or electron pairs will cohere to form superconductors.  
For example, in two-dimensional sheet of cuprate superconductivity, it is less 
than $0.6/3^2=0.067$, where $0.6$ is threshold of square in site percolation, 
$3$ is theoretical minimum for radius of delocalization divided by one of 
localization, and $2$ comes from dimensionality. This threshold is compare with 
our critical fraction $0.05$ in phase diagram of cuprate conductors. 

 In three dimensional metals, for example in face-centred cubic(FCC) copper, the 
 threshold of (super)conductivity is less than $0.2/5^3=0.0016$, where $0.2$ is 
 threshold of site percolation for FCC lattice, $5$ is minimum for radius of 
 delocalization divided by one of localization in Fig. 1b, and $3$ comes from 
 dimensionality. It is compare with $0.0001$\cite{Moura}, fraction of charge 
 carriers in metal superconductivity. From above instances we knew that 
 thresholds of  percolation of (super)conductivity were very sensitive to the 
 dimensionality of (super)conductors and were very small in three dimensional 
 case so that electron pairing and its condensation of coherence occur almost 
 simultaneously, while the critical fraction in two dimensional sheet of cuprate 
superconductor is rather large so that electrons pairing should occur before its 
 condensation of coherence distinctly. So we predict that superconductors at 
 room temperature should be ionic compound of delocalized $s$ electrons because 
 they should have high barrier between delocalization and localization and small 
thresholds of percolation in three dimension, which may find from the other end 
 of $3d$ transition metal oxide such as Sc$_2$O$_{3-x}$, TiO$_{2-x}$ or 
 V$_2$O$_{5-x}$,  because their $4s$ and $3d$ orbitals are both near Fermi 
 surface to form $s$ type conductors when insulators transit to conductors.


\section{conclusion}
\label{conc}


We found the localization-delocalization duality of $s$ electrons in metals, 
from which we unified the theory of conductivity for metals, Mott insulators and 
organic conductors. The phase diagram of cuprate superconductors also be 
explained by ideal Mott insulator-metal transition qualitatively. We also 
predicted that higher temperature superconductor might be ionic compound with 
delocalized $s$ electrons.



\end{document}